\documentclass[pra,twocolumn,showpacs,superscriptaddress,floatfix]{revtex4-2}
\usepackage{graphicx}
\usepackage{epstopdf} 
\usepackage{amsmath}
\usepackage{epsfig}
\usepackage{latexsym}
\usepackage{amsfonts}
\usepackage{graphics}
\usepackage{bm}
\usepackage{braket}
\usepackage{dsfont}
\usepackage{soul}
\usepackage{ulem}
\usepackage{bbm}
\usepackage{hyperref}

\usepackage{mathtools}

\usepackage{helvet}

\graphicspath{ {./Graphics/} }
\begin{document}

\date{\today}
\author{B. Lajci}
\affiliation{Department of Physics and Astronomy, McMaster University, 1280 Main St.\ W., Hamilton, ON, L8S 4M1, Canada}
\affiliation{Perimeter Institute for Theoretical Physics, Waterloo, ON, Canada N2L 2Y5}
\author{D. H. J. O'Dell}
\affiliation{Department of Physics and Astronomy, McMaster University, 1280 Main St.\ W., Hamilton, ON, L8S 4M1, Canada}
\author{J. Mumford}
\affiliation{Department of Physics and Astronomy, McMaster University, 1280 Main St.\ W., Hamilton, ON, L8S 4M1, Canada}
\affiliation{Homer L. Dodge Department of Physics and Astronomy, The University of Oklahoma, Norman, Oklahoma 73019, USA}
\affiliation{Center for Quantum Research and Technology, The University of Oklahoma, Norman, Oklahoma 73019, USA}
\affiliation{Department of Physics and Astronomy, University of Victoria, Victoria, British Columbia V8P 5C2, Canada}

\title{Topologically protected Bell-cat states in a simple spin model}

\begin{abstract}
We consider the topological properties of the so-called central spin model that consists of $N$ identical spins coupled to a single distinguishable central spin which arises in physical systems such as circuit-QED and bosonic Josephson junctions coupled to an impurity atom. The model closely corresponds to the Su-Schrieffer-Heeger (SSH) model except that the chain of sites in the SSH model is replaced by a chain of states in Fock space specifying the magnetization. We find that the model accommodates topologically protected eigenstates that are `Bell-cat' states consisting of a Schr\"{o}dinger cat state of the $N$ spins that is maximally entangled with the central spin, and show how this state can be adiabatically created and moved along the chain by driving the central spin.  The Bell-cat states are visualized by plotting their Wigner function and we explore their robustness against random noise by solving the master equation for the density matrix. We also explain the essential topological difference between identical spins  and the excitations of a bosonic mode.
\end{abstract}

\maketitle

\section{\label{Sec:Intro}Introduction}

A cat state is a coherent superposition of two or more quantum states that exhibit significant differences on a macroscopic scale. Initially introduced in a thought experiment by Erwin Schrödinger \cite{schrodinger35} to underscore the striking distinctions between quantum and classical physics, cat states became a useful concept in philosophical discussions about the interpretations of quantum mechanics.   More recently, progress in controlling quantum systems at the level of single atoms and photons has led to the successful realization of cat states in experiments. These include the superposition of a single laser cooled ion in two spatially separated locations \cite{monroe96}, and the superposition of two or more coherent states of photons localized in different regions of phase space in a light pulse   \cite{ourjoumtsev07}, in microwave fields in cavities  \cite{deleglise08}, and in circuit-QED systems \cite{vlastakis13}. Additionally, cat states have been created using the spin degrees of freedom of trapped ions \cite{leibfried05}, and also by placing the vibrational state of a trapped ion in a coherent superposition dependent on its internal spin \cite{wineland13,lo15,kienzler16}. Such entanglement between different types of degrees of freedom has also been achieved between the polarization of spatial mode degrees of freedom with photons  \cite{gao10}. 

Progress in our ability to generate and control quantum states has technological implications because cat states formed from a pair of coherent states have the potential to be used as qubits \cite{haroche06}.  When sufficiently separated, each coherent state has exponentially small overlap, so they can form the basis states $\vert 0 \rangle$ and $\vert 1 \rangle$ of the qubit.  Coherent states have significantly larger Hilbert spaces than traditional qubits which allows for the possibility of encoding  more information within these superpositions, and in some cases, facilitating more effective error correction protocols \cite{rosenblum18,grimm20,lescanne20}. 

A widely implemented technique for creating cat states involves the use of an ancillary qubit. The first step is to initialize the qubit in an equal superposition of its two states, which for an atomic qubit would be achieved by applying a $\pi/2$ laser pulse.  The second step is to entangle the qubit with a coherent state of a larger quantum system, thereby generating a cat state in the latter system. This can be accomplished by making the phase of the coherent state dependent on the state of the qubit.  For example,  a force applied to a mechanical oscillator that depends on the state of a qubit  can generate a macroscopic superposition of two coherent states of phonons where each coherent state has a different phase  \cite{bild23}, or the state of an atom resonantly coupled to an optical cavity can be used to apply a phase flip to incoming coherent pulses of light  \cite{hacker19}. In numerous instances, the interaction between the ancillary qubit and the coherent state results in maximal entanglement between the two, giving rise to states known as Bell-cat (BC) states, see \cite{vlastakis15} and references therein (these states are also known by other names such as squeezed wavepacket entangled states \cite{lo15}). BC states represent more complex counterparts to regular Bell states, which in their original setting are maximally entangled states only of pairs of qubits. The complexity arises from the fact that coherent states possess more degrees of freedom than a single qubit, and this has lead to BC states being employed in research exploring information extraction and entanglement generation beyond the scope of qubit Bell states \cite{vlastakis15}. The continuous nature of the relative phase between coherent states means BC states might find a role in the continuous variable form of quantum information processing \cite{braunstein15}.

In this paper, we show theoretically that a BC state can be created via adiabatic driving in a simple version of the Mermin central spin (CS) model \cite{hutton04,breuer04,alhassanieh06,garmon11}. The CS model consists of a central two-state particle, which plays the role of an ancillary qubit, coupled to $N$ identical, non-self-interacting two-state particles. The driving protocol leverages two key features of this system: (1) the presence of a topological phase transition, where the topologically nontrivial phase supports a pair of zero-energy bound states in the Fock space of the identical spins; and (2) the inhomogeneous structure of the Fock space, which allows the locations of the bound states to be tuned by varying the strength of an applied magnetic field.  Each bound state is coupled to a different state of the central spin, and because they are degenerate, they can be made to form a zero-energy BC state.  The eigenstates that will become the BC state in the topologically nontrivial phase are localized at the same location in Fock space in the trivial phase.  To prepare the BC state, we start with an unentangled initial state that has substantial overlap with these localized eigenstates in the trivial phase. By slowly tuning the applied magnetic field through the topological phase transition, the state evolves adiabatically into a BC state, where the ancillary qubit becomes maximally entangled with the ensemble of $N$ identical particles.

We analytically confirm that the pair of bound states are well localized in Fock space and closely resemble coherent spin states. Numerical results further demonstrate that these states are separated from the bulk of the spectrum by a relatively large energy gap. As in condensed matter systems, the bound states are protected by an underlying symmetry of the Hamiltonian, making them robust against symmetry-preserving perturbations. These three properties make them ideal target states for the driving protocol. While the resulting BC state remains susceptible to decoherence, particularly from random noise in the driven magnetic field, we find that the state forms rapidly upon entering the topologically nontrivial phase. This suggests that, in practice, the BC state can be realized within a timescale short enough to outpace decoherence.

Highly controllable versions of the CS model can potentially be realized experimentally using the physical systems already mentioned. For example, the two polarization states of photons means they can naturally play the role of the non-self-interacting two-state particles. Alternatively, the two topologically distinguished clockwise and anticlockwise orbital angular momentum states of photons in a structured Gaussian light beam can also act in this capacity \cite{dennis17,gutierrez20}. Coupling photons to a real atom \cite{deleglise08}, or a circuit-QED setup in the form of a superconducting transmon qubit in a microwave waveguide cavity resonator  \cite{vlastakis13,vlastakis15}, would allow the realization of the CS model, as would the spin-motion entanglement that can be utilized in trapped ion systems \cite{lo15}. 

Yet another way to realize the CS model is using ultracold atoms, for example by immersing an atomic quantum dot in a two-state Bose-Einstein condensate (BEC), the latter also being known as a bosonic Josephson junction. The two states of the BEC atoms can either be two internal spin states \cite{zibold10} or two external modes (double well potential) \cite{albiez05,levy07}, and the atomic quantum dot can be a single impurity atom \cite{bausmerth07,rinck11,mulansky11,mumford14a,mumford14b,chen21} or ion \cite{gerritsma12,joger14,ebgha19} that also has either two internal or two external states. Feshbach resonances can be used to independently control the boson-boson interactions and impurity-boson interactions. Recent experimental realizations of the controlled coupling between individual localized Cs impurity atoms immersed in a Rb BEC in have been reported in references \cite{schmidt18,bouton20,adam22}.We also note that bosonic Josephson junctions can also be realized using structured light in an aberrated optical cavity \cite{gutierrez23}.

\section{\label{Sec:Mod}Model}

The system we will be investigating is a version of the CS model which describes $N$ identical two-state particles interacting with a distinguishable central two-state particle.  Due to each particle having access to two states, we will refer to them as spin-1/2 particles.  The $N$ identical spins do not interact with each other, but interact equally with the CS leading to the Hamiltonian    

\begin{eqnarray}
\hat{H} = v \hat{\sigma}_x + \frac{2w}{N}\left ( \hat{S}_+ \hat{\sigma}_- + \hat{S}_- \hat{\sigma}_+ \right )
\label{eq:ham}
\end{eqnarray}
where the explicit Pauli matrices operate on the CS, whereas the large collective spin operators $\hat{S}_\pm = \sum_i^N \hat{\sigma}_\pm^i$  act on the $N$ identical spins. 
 The parameters $v$ and $w$ represent the CS spin-flip energy and the energy of the spin-exchange between the CS and identical spins, respectively.  We will work in the eigenbasis  of the $\hat{S}_z$ and $\hat{\sigma}_z$ operators, $\{\vert n, m \rangle \}$, where $n = (N_\uparrow - N_\downarrow)/2$ is half the difference between the number of up and down identical spins and $m = \uparrow, \downarrow$ are the two states of the CS. This basis defines the Fock states of the system and physically speaking it gives the magnetization along the $z$ axis.

 The usefulness of this  basis comes from the fact  that the CS model bears a close resemblance to the Su-Schrieffer-Heeger (SSH) model \cite{mumford23} which describes the electronic properties of a 1D polymer with alternating single and double bonds.  Under this comparison the Fock state label $n$ is analogous to the unit cell label and $m$ is analogous to the intra-unit cell label.  In addition, like the SSH model, the CS model undergoes a topological phase transition from a trivial phase when $v>w$ to a nontrivial phase when $v<w$ which is marked by the presence of a pair of topologically protected bound states with energy $E = 0$.  However, unlike the SSH model whose bound states only appear at the edges of the polymer, the CS model's bound states can be centered at any value of $n$.  Their protection relies on the chiral symmetry which the Hamiltonian in Eq.\ \eqref{eq:ham} obeys and is  defined in terms of the chiral operator $\hat{\sigma}_z$ and the relation $\hat{\sigma}_z\hat{H}\hat{\sigma}_z = - \hat{H}$.  The chiral symmetry means that $\hat{\sigma}_z$ transforms states with energy $E$ to states with energy $-E$ and since the bound states have $E=0$, they are eigenstates of  $\hat{\sigma}_z$.  This also means that each bound state exists entirely in either the CS $\uparrow$ or $\downarrow$ subspace and that any perturbations respecting the chiral symmetry will not destroy them.

In condensed matter systems, topologically protected bound states are well localized in the material.  For now we will assume that the same is true for the protected bound states in the Fock space of the CS model and provide evidence for it later.  The assumption of localization allows us to make a mean-field approximation for the identical spins' operators by taking their coherent spin expectation values

\begin{equation}
\langle \theta, \phi \vert \bm{\hat{S}} \vert \theta, \phi \rangle = \frac{N}{2} \left ( \sin\theta\cos\phi, \sin\theta\sin\phi, \cos\theta \right ).
\end{equation}
This approximation labels the states of the identical spins as points on the surface of a spin-$N/2$ Bloch sphere where $\theta$ and $\phi$ are the polar and azimuthal angles, respectively.  The mean-field approximation of Eq.\ \eqref{eq:ham} is 

\begin{equation}
\mathcal{H}(\theta, \phi) =\bm{d}(\theta, \phi) \cdot \bm{\hat{\sigma}} 
\label{eq:bloch}
\end{equation}
where $\bm{\hat{\sigma}} = (\hat{\sigma}_x,\hat{\sigma}_y,\hat{\sigma}_z)$ and the vector $\bm{d}(\theta,\phi)= \left ( d_x(\theta,\phi), d_y(\theta,\phi),d_z(\theta,\phi)\right)$ has components 

\begin{eqnarray}
d_x(\theta, \phi) &=& v+w\sin\theta\cos\phi, \hspace{10pt} d_y(\theta, \phi) = w\sin\theta\sin\phi,  \nonumber \\
d_z(\theta, \phi) &=& 0.
\label{eq:comp}
\end{eqnarray}
Equation \eqref{eq:bloch} resembles a Bloch Hamiltonian of a two-band model describing a periodic lattice under the tight-binding approximation, however, there is a major difference.  In a typical Bloch Hamiltonian the vector $\bm{d}$ is parameterized by the quasi-momentum which is $2\pi$-periodic, so it maps points from a circle (1D) or a torus (2D) to, in general, a 3D vector, whereas $\bm{d}$ in Eq.\ \eqref{eq:bloch} maps points from the mean-field Bloch sphere to a 2D vector ($d_{z}=0$).   The azimuthal angle $\phi$ is indeed $2\pi$-periodic, so it plays the role of a quasi-momentum in the CS model, but the polar angle $\theta$ is not $2\pi$-periodic, so it cannot be considered as another quasi-momentum.  Instead, it gives the `position' in Fock space from the relation $n = \frac{N}{2} \cos\theta$.  

In 1D lattice models which undergo a topological phase transition, the transition is quantified in terms of the winding number which answers the question, how many times does $\bm{d}$ wind around its origin as the quasi-momentum of occupied bands is cycled through its range once.  Due to the fact that $\bm{d}$ maps one periodic surface to another, the winding number takes integer values and is nonzero in the topologically nontrivial phase.   In the SSH model, the components of $\bm{d}$ can be written in a similar form to Eq.\ \eqref{eq:comp},  but with  $\theta$ fixed to $\theta= \pi/2$ and $\phi \to k$ where $k$ is the quasimomentum of the lattice:
\begin{eqnarray}
d^{\mathrm{SSH}}_x(k) &=& v+w\cos k, \hspace{10pt} d^{\mathrm{SSH}}_y(k) = w \sin k,  \nonumber \\
d^{\mathrm{SSH}}_z(k) &=& 0.
\label{eq:compSSH}
\end{eqnarray}
Therefore, as $k$ varies by $2\pi$, $\bm{d}$ traces out a circle of radius $w$ in the $d_x$, $d_y$ plane centered at $(v,0)$. When $v<w$, the circle encloses the origin and the winding number is unity which signifies that the system is in the topologically nontrivial phase.  The components of $\bm{d}$ for the CS model are more general than those of the SSH model, however, we can ask a similar question: how many times does $\bm{d}$ wind around the origin of the $d_x$, $d_y$ plane for \textit{any} fixed value of $\theta$ as $\phi$ varies by $2\pi$. We answer this question with an explicit calculation of the winding number 

\begin{eqnarray}
W(\theta) &=& \frac{1}{2\pi} \int_0^{2\pi} \left ( \bm{\hat{d}}(\theta,\phi) \times \frac{d}{d\phi}\bm{\hat{d}}(\theta,\phi)\right )_z d\phi \nonumber \\
&=& \frac{1}{2} \left [ 1+\mathrm{sgn}\left (w\sin\theta-v \right ) \right ]
\label{eq:wind}
\end{eqnarray}
where $\bm{\hat{d}} = \bm{d}/\vert \bm{d}\vert$.  The topologically trivial phase is marked by $W=0$ which occurs when $v>w$ and the topologically nontrivial phase is marked by $W = 1$ which occurs when $v<w$.  However, due to the $\theta$ dependence of the winding number, only part of the Bloch sphere of the identical spins has nontrivial topology in the latter case.  The nontrivial part lies in the range $\theta_1 < \theta < \theta_2$ where $\theta_1 = \arcsin\left(v/w\right)$ and  $\theta_2 =\pi -  \arcsin\left(v/w\right)$ and outside of this range there is trivial topology.  The boundary angles mark the polar coordinates on the Bloch sphere of topologically protected bound states \cite{mumford23} and we emphasize that the appearance of a single bound state at each boundary provides support for the bulk-boundary correspondence. The correspondence asserts that a bulk topological invariant, such as the winding number, determines the number of bound states at a particular boundary separating two regions of different topology. In the CS model, the presence of two boundaries is a consequence of the existence of two `edges' in the Fock space of the identical spins located at $n_\mathrm{edge} = \pm N/2$.

\begin{figure}[t]
\centering
\includegraphics[scale=0.5]{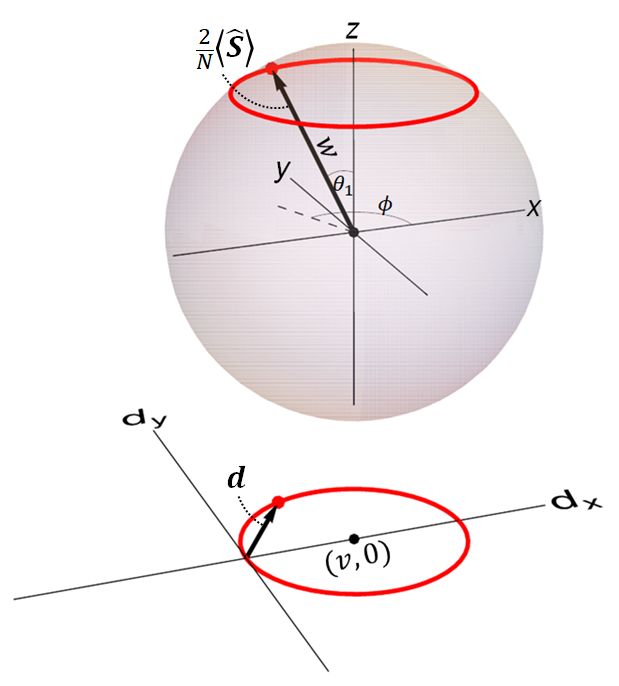}
\caption{Geometric representation of the mean-field Hamiltonian.  The vector pointing from the origin of the sphere to its surface represents the mean-field state of the identical spins while the direction of the vector $\bm{d}$ in  the $d_x$, $d_y$ plane gives the eigenstate of the mean-field Hamiltonian.  For a fixed value of $\theta$, as $\phi$ changes by $2\pi$, the sphere vector traces out a circuit (red).  The vector $\bm{d}$ follows the same path, but projected directly down onto the $d_x$, $d_y$ plane.  When the projected circuit encompasses the origin of the plane, the system is in the topologically nontrivial phase with winding number $W=1$.  The red circle at $\theta = \theta_1$ is the boundary circuit between the topologically trivial and nontrivial phases where any circle on the sphere with a larger radius will encompass the origin when projected down onto the plane.}
\label{fig:WB}
\end{figure}

Figure \ref{fig:WB} gives a geometric interpretation of the topological properties of the system.  The vector extending from the sphere's center to its surface is $\frac{2}{N}\langle \bm{\hat{S}}\rangle$ (scaled by $w$) and represents the mean-field state of the identical spins while the vector in the $d_x$, $d_y$ plane is $\bm{d}(\theta,\phi)$ whose direction denotes the eigenstate of the mean-field Hamiltonian in Eq.\ \eqref{eq:bloch}.  For a fixed value of $\theta$, as $\phi$ changes by $2 \pi$, the vector on the sphere traces out a circle on its surface and the vector $\bm{d}(\theta,\phi)$ traces the same path projected directly down onto the $d_x$, $d_y$ plane.  The $d_x$ coordinate of the center of the sphere is $v$, whereas the sphere's radius is $w$.  When $v > w$, the projected path fails to enclose the origin of the plane, causing $\bm{d}(\theta,\phi)$ to oscillate without fully winding as $\phi$ changes by $2\pi$.  When $v<w$, there exist certain values of $\theta$ for which the trajectory projected onto the plane encircles the origin, causing $\bm{d}(\theta,\phi)$ to wrap around the origin once.  The range of $\theta$ exhibiting nonzero winding is  $\theta_1 < \theta < \theta_2$, with the boundaries determined by the points where the projected trajectory intersects the origin of the plane.  In Figure \ref{fig:WB}, the red circles represent the trajectories for $\theta = \theta_1$ and show that when $\phi = \pi$,  $\vert \bm{d}(\theta_1,\pi) \vert = 0$ leading to the intersection of the trajectory with the origin and the closure of the energy gap, $E_{\mathrm{gap}} = 2 \vert \bm{d} \vert$, at that point.   Although not shown, due to the spherical symmetry, the same thing happens at $(\theta, \phi) = (\theta_2, \pi)$.  Consequently, these points correspond to the locations of the protected bound states.  

Figure \ref{fig:WB} also gives a clear picture of how modifications to Eq.\ \eqref{eq:ham} affect the topological properties of the system.  For instance, if the term $u \hat{\sigma}_y$ is introduced, then the center of the sphere will have its coordinates shifted to $(v,u)$ in the plane.  The condition for the existence of a nontrivial topological region is that the radius of the sphere be larger than the distance from the origin to its center projected onto the $d_x$, $d_y$ plane which means the topological phase transition will occur at $w= \sqrt{v^2+ u^2}$.  Having nonzero $u$ also changes the direction from the origin to the sphere, so the gap will no longer close at $\phi = \pi$, but at the more general angle $\phi = \pi + \arctan(u/v)$.  Note that such a  change preserves the chiral symmetry of the Hamiltonian and in turn preserves the existence of protected bound states.  Modifications that introduce terms proportional to $\hat{\sigma}_z$ break the chiral symmetry by lifting the projected trajectories out of the $d_x$, $d_y$ plane, so they are no longer forced to intersect the origin.  In that case, the energy gap no longer closes and the protected bound states no longer exist.

\section{\label{Sec:Res}Results}

\subsection{Bound state fluctuations}
Before describing our protocol for the the creation of BC states we will confirm the existence of the protected bound states discussed in the previous section.  The mean-field analysis determined the locations of the bound states, however, we would also like to know about their fluctuations to determine how localized they are.  For that, we use the fact that the bound states are zero energy eigenstates and exist entirely in one of the CS subspaces, so $\mathcal{H}(n,\phi) \chi_{\uparrow(\downarrow)}(n) = 0$ where 

\begin{eqnarray}
 \chi_{\uparrow}(n) &=& B_\uparrow(n) 
 \begin{bmatrix}
         1 \\
	0
 \end{bmatrix} \nonumber \\
\chi_{\downarrow}(n) &=& B_\downarrow(n)
 \begin{bmatrix}
         0 \\
	1
 \end{bmatrix}.
\end{eqnarray}
The bound states are written as products of the wave function in the identical spins' Fock space $B_{\uparrow(\downarrow)}(n)$ and a two component vector for the two states of the spin-1/2 particle.  Note that we have changed from the polar coordinate, $\theta$, to the $\hat{S}_z$ state label, $n$, through the relation $n = \frac{N}{2} \cos\theta$ which will be useful in the coming calculations.  In the $n$ state space the bound states are located at $n_1 = n_b$ and  $n_2 = -n_b$ where $n_b = \frac{N}{2}\sqrt{1-\left (\frac{v}{w} \right )^2}$.  We are  interested in fluctuations around  the gap closing point at $\phi = \pi$, so we expand  $\mathcal{H}(n,\phi)$ to linear order around this point.  In addition, we incorporate fluctuations into the mean-field approximation by making use of the fact that the number variable $n$ and the phase variable $\phi$ form a conjugate pair \cite{josephson62,raghavan99} so the transformation $\phi \to i \frac{\partial}{\partial n}$ can be made resulting in the differential equation

\begin{eqnarray}
 &&\left [\left ( v - w\sqrt{1-\left ( \frac{2n}{N}\right)^2} \right ) \hat{\sigma}_x \right.\nonumber \\
&& \left.-w\sqrt{1-\left ( \frac{2n}{N}\right)^2} \left (i \frac{\partial}{\partial n} - \pi \right ) \hat{\sigma}_y \right ]\chi_{\uparrow(\downarrow)} (n) =0  \ .
\label{eq:e0}
\end{eqnarray}
The wave functions $B_{\uparrow(\downarrow)}(n)$ can be solved for exactly and are of the form $B_\uparrow(n) \propto e^{i \pi n}e^{n-\frac{Nv}{2w} \arcsin\left ( \frac{2n}{N}\right )}$ where $ B_\downarrow(n) =B_\uparrow (-n)$, however, they are only valid in the region around $\pm n_b$, so we expand them around those points giving 

\begin{equation}
B_{\uparrow(\downarrow)}(n) \approx \left ( 2\pi\sigma^2\right )^{-1/4} e^{i\pi n}e^{-\frac{\left (n \pm n_b \right )^2}{4 \sigma^2}}
\label{eq:EBS}
\end{equation}
where $\sigma = \sqrt{\frac{N v^2}{4w^2\sqrt{1-\left( \frac{v}{w}\right )^2}}}$ is the standard deviation of the wave function.  The bound states take the form of Gaussians centered at the gap closing locations with widths that scale as $1/\sqrt{N}$ relative to $N$, so for a sufficiently large number of identical spins, they are well localized.  The presence of the $\left [1-(v/w)^2\right ]^{-1/2}$ factor in $\sigma$ also indicates that the bound states become increasingly localized as the system moves deeper into the topologically nontrivial phase.  Indeed, in the limit $v/w \to 0$ the bound states are Fock states located at $\pm N/2$.  As shown in Figure \ref{fig:WB}, in this limit, the Bloch sphere sits directly above the origin of the plane, resulting in the bound states being positioned at the north and south poles of the Bloch sphere.

\begin{figure}[t]
\centering
\includegraphics[scale=0.15]{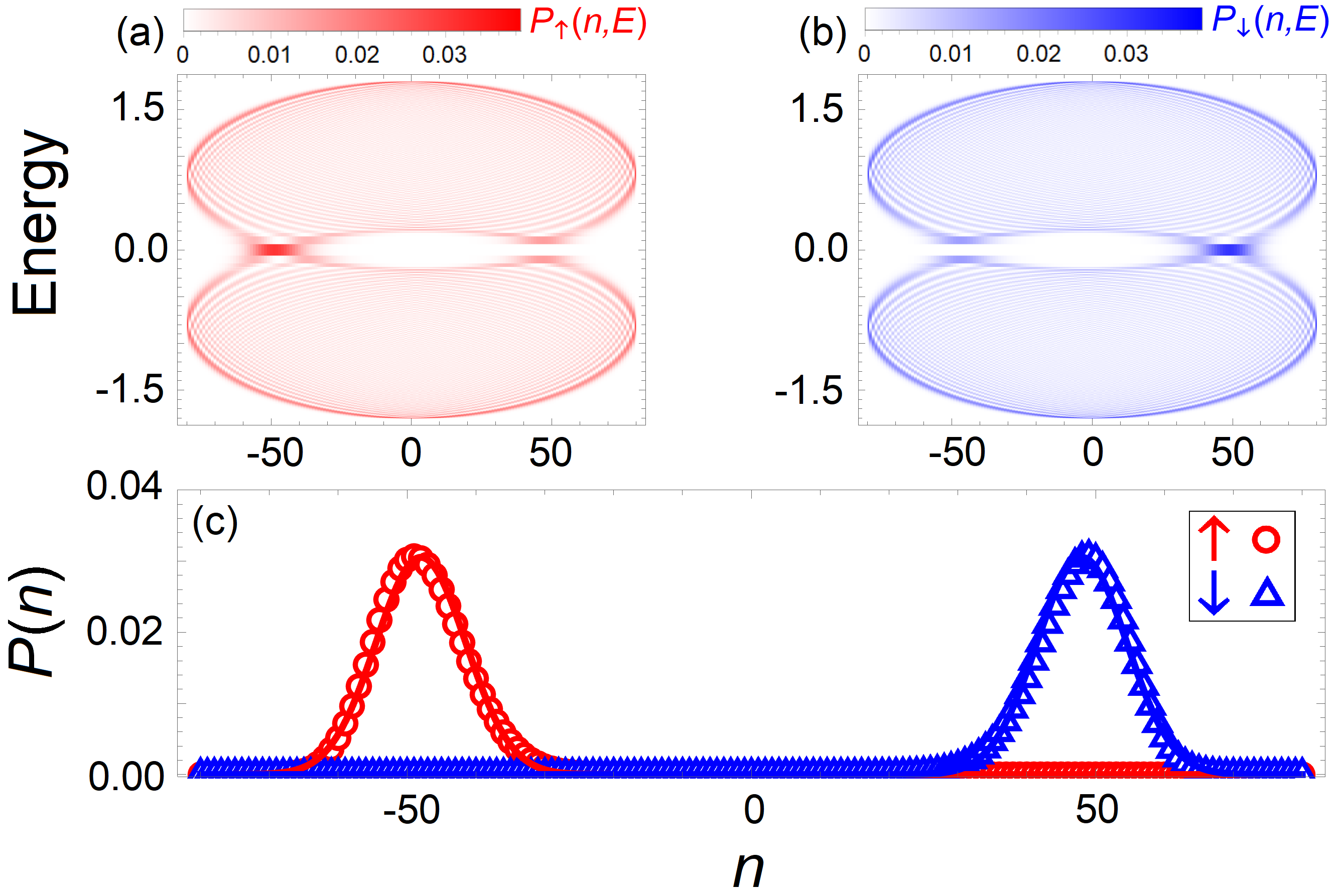}
\caption{Probability distributions in the Fock and energy bases.  (a) Probability for a Fock state to have a given energy in the CS $\uparrow$ subspace,  $P_\uparrow (n,E) = \vert \langle E \vert n,\uparrow \rangle \vert^2$.  (b) Same as (a), but in the CS $\downarrow$ subspace, $P_\downarrow (n,E) = \vert \langle E \vert n,\downarrow \rangle \vert^2$.  (c) Probability distributions of the zero energy bound states where the left and right distributions are in the CS $\uparrow$ (red circles) and $\downarrow$ (blue triangles) subspaces, respectively.  The parameter values are $N = 180$, $w = 1$ and $v = 0.7$.}
\label{fig:RB}
\end{figure}

Figure \ref{fig:RB}(a) shows $P_\uparrow (n,E) = \vert \langle E \vert n,\uparrow \rangle \vert^2$ which is  the probability for a Fock state to have energy $E$ in the CS $\uparrow$ subspace.  Two lobes are shown at positive and negative energies with a single bound state between them at $E = 0$.  Panel (b) shows the same figure except in the CS $\downarrow$ subspace where the only difference is that the bound state is located on the other side of $n = 0$.  Panel (c) shows the $E = 0$ slice of both (a) and (b), $P(n) =\vert \langle E = 0\vert n, \uparrow \rangle + \langle E = 0 \vert n, \downarrow \rangle \vert^2/2$, where the red circles and blue triangles are data from the diagonalization of Eq.\ \eqref{eq:ham}, while the solid lines are $\vert B_\uparrow(n)\vert^2$ and  $\vert B_\downarrow(n)\vert^2$ from Eq.\ \eqref{eq:EBS}, respectively.  There is excellent agreement between the analytic and numerical results, and this is expected to hold as long as the system remains away from the topological phase transition point at $v=w$. This is because, as previously mentioned, the width of the bound states is predicted to diverge at this point, leading to discrepancies between the analytic and numerical results due to finite size effects.

\subsection{Generating BC states}

The protocol we use to generate BC states can be broken down into three steps.  (1) The system is initialized in the topologically trivial phase ($v>w$) in the product state 

\begin{equation}
\vert\psi(0)\rangle = \vert \pi/2,\pi \rangle \otimes \left ( \vert \uparrow \rangle + \vert \downarrow \rangle\right )/\sqrt{2}
\label{eq:init}
\end{equation}
where $\vert \pi/2, \pi \rangle$ is the spin coherent state centered at $\left ( \theta, \phi \right ) = \left (\pi/2,\pi \right )$ on the Bloch sphere.  For large $N$, the coherent part has a Gaussian shape centered at $n = 0$ in the $\hat{S}_z$ basis 

\begin{equation}
\langle n \vert\pi/2,\pi\rangle \approx (\pi N/2)^{-1/4} e^{i\pi n}e^{-\frac{n^2}{N}}.
\label{eq:coh}
\end{equation}
Our motivation for starting the process in the trivial phase comes from the fact that eigenstates of the Hamiltonian which will become the topologically protected  bound states are centered at $\theta = \pi/2$ ($n = 0$ in Fock space) and $\phi = \pi$ on the Bloch sphere in the trivial phase.  Furthermore, we start with an equal superposition of the two CS states because we recall that in the nontrivial phase each of the two bound  states is in opposite subspaces of the CS.  (2) Next, we adiabatically drive the system into the topologically nontrivial phase by linearly ramping down an applied magnetic field which controls the CS spin-flip energy, 

\begin{equation}
v(t) = v_0 - \gamma t,
\end{equation}
where $v_0$ is the initial value of $v$ and $\gamma$ is the drive rate.  Because the initial state is in a superposition of the CS being both up and down, the wave function will split in two with each part being attached to $B_\uparrow(n)$ and $B_\downarrow(n)$.  (3)  After some time when the two parts of the wave function have negligible overlap, they are in a macroscopic superposition of the majority of the identical spins pointing up and pointing down.  The splitting process relies on the fact that the topological bound states are protected by the chiral symmetry and are therefore eigenstates of $\hat{\sigma}_z$.  This forces the bound states on opposite sides of $n = 0$ to have support in opposite CS subspaces resulting in the formation of a BC state.

\begin{figure}[t]
\centering
\includegraphics[width=\columnwidth]{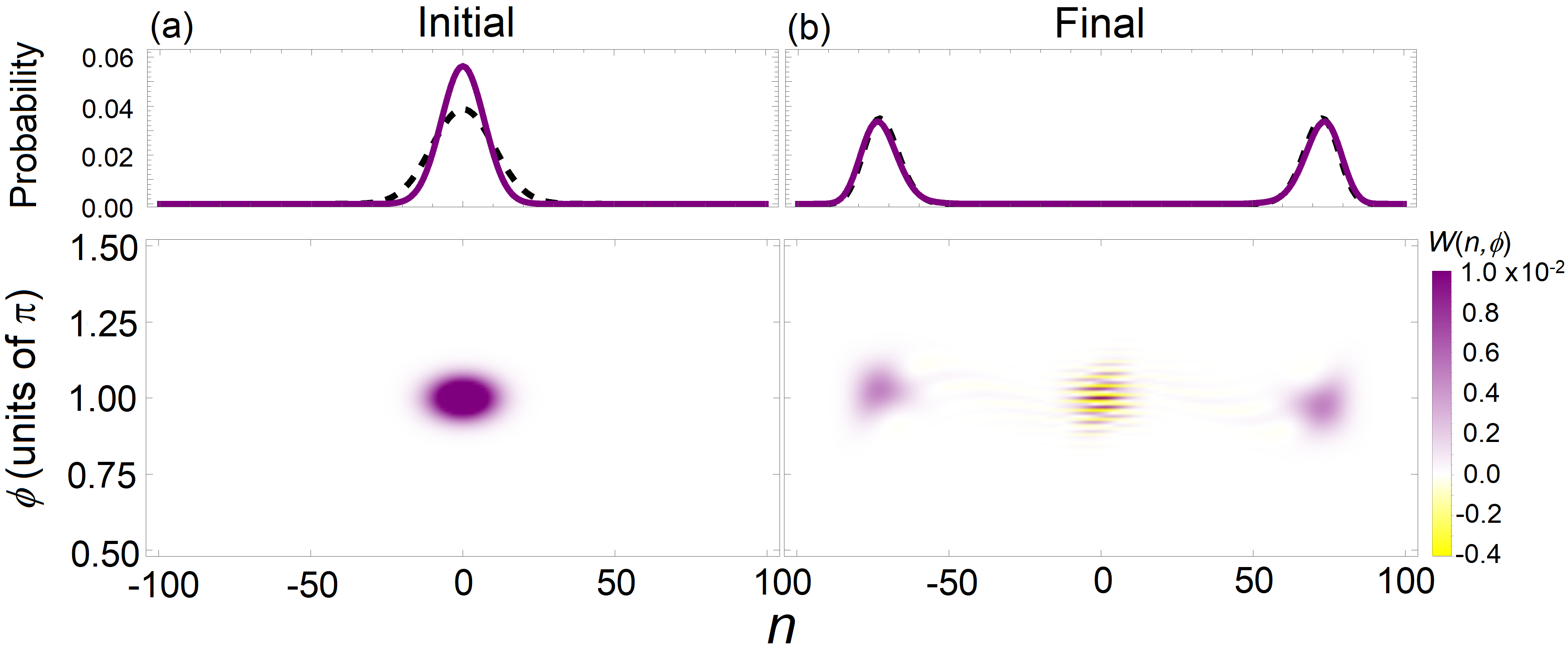}
\caption{Wigner quasi-probability distributions. (a) The bottom panel shows the sum of the zero and $x$ components of the joint Wigner quasi-probability distribution of the spin coherent state $\vert \pi/2, \pi \rangle$.  The top panel shows the probability distribution of the coherent state in Fock space (solid purple) and the even parity bound state (dashed black) in the topologically trivial phase where $v = 1.3$.    (b) The top and bottom panels show the same as (a), but for the final state at the end of the driving period.  The parameters are $\gamma \approx 1.2 \times 10^{-3}$, $v_0 = 1.3$ and $v_f = 0.7$.   The dashed black curve in the top panel is the even parity bound state in the topologically nontrivial phase where $v = 0.7$.   In all panels $w = 1$ and $N = 200$.}
\label{fig:wig}
\end{figure}

To illustrate the initial and final steps of the splitting process we calculate the joint Wigner quasi-probability distribution

\begin{equation}
W_i (n,\phi_p) = \frac{1}{N+1} \sum_{n^\prime} \langle n-n^\prime \vert \rho_i \vert n + n^\prime \rangle e^{2i \phi_p n^\prime}.
\end{equation}
which gives the state of the system in the space of conjugate variables $(n,\phi)$ and is similar to the phase space of classical systems.  The phase $\phi_p = \frac{2\pi p}{N+1}$ is the discrete version of the azimuthal angle of the identical spins' Bloch sphere and $\rho_i = \mathrm{Tr}_\mathrm{CS}\left [\hat{\sigma}_i \rho\right ]$ where $i = 0,x,y,z$. In experiments, a complete representation of a given state requires all four distributions.  However, for visualizing both the distribution splitting and quantum correlations, only $W_0(n, \phi_p)$ and $W_x(n, \phi_p)$ are necessary. Both are combined in Figure \ref{fig:wig}, where the lower part of (a) illustrates the initial state with $\rho(0) = \vert \psi(0)\rangle \langle \psi(0) \vert$, and the lower part of (b) depicts the final state once the driving has stopped.  

The central region of the lower part in (b) displays negative values originating from $W_x(n, \phi_p)$, a distinctive feature reflecting quantum correlations in the Wigner distribution.  The two blobs on opposite sides of $n = 0$ represent the macroscopic superposition.  Together, these distributions clearly indicate the formation of a cat state.  The solid purple curve in the top of both (a) and (b) shows $\sum_p W_0(n,\phi_p)$ which is the probability distribution in the identical spins' Fock space. The dashed black curves are the probability distributions of the bound states, representing the target states.  There is excellent overlap between the evolved and target distributions following the driving in (b).

\begin{figure}[t]
\centering
\includegraphics[scale=0.16]{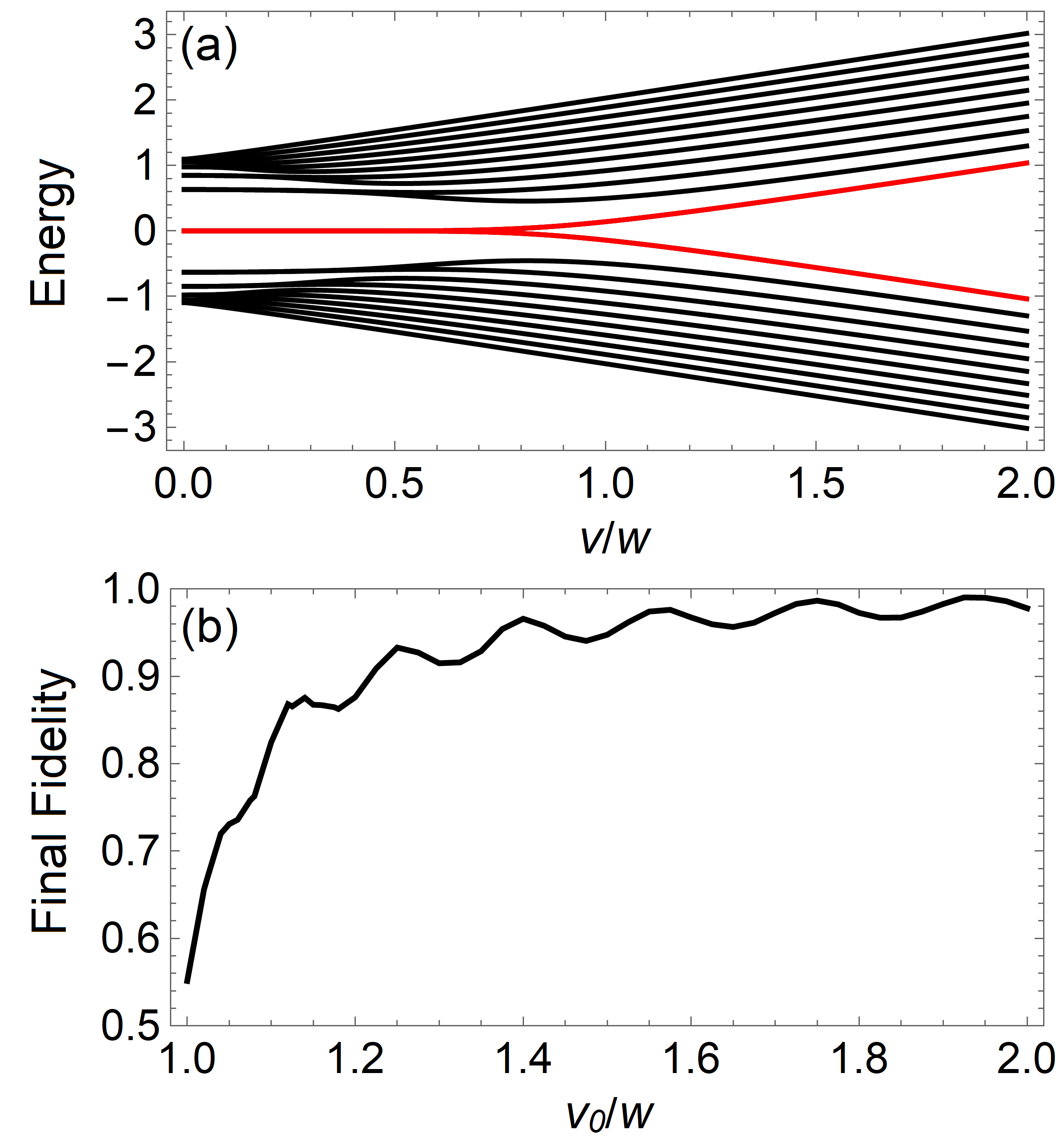}
\caption{Energy spectrum and final state fidelity.  (a) Energy spectrum as a function of the scaled CS spin-flip energy $v/w$.  The red and black curves represent the energies of the bound and bulk states, respectively.  The parameters are $N = 10$ and $w = 1$.  (b) Fidelity between the final state at the end of the driving period and the BC state as a function of the scaled initial CS spin-flip energy $v_0/w$.  The parameters are $N = 150$, $w = 1$, $\gamma \approx 1.2 \times 10^{-3}$ and $v_f = 0.7$. }
\label{fig:fid}
\end{figure}

Two primary factors influencing the quality of the final BC state are adiabaticity and the overlap between the initial coherent state and the energy eigenstate which will become the bound states.  To ensure adiabaticity, we require that the drive rate is small enough to satisfy \cite{amin09},

\begin{equation}
\max\limits_{v_0 \leq v \leq v_f} \frac{\vert \langle E_n (v) \vert \dot{\hat{H}} \vert E_m(v) \rangle \vert}{\vert E_n(v) - E_m(v) \vert^2} \ll 1, \hspace{15pt} \text{for all} \hspace{5pt} m \neq n 
\label{eq:adia}
\end{equation}
%
where $v_f$ is the final value of $v$ and $\vert E_n(v)\rangle$ is the $n^\mathrm{th}$ instantaneous energy eigenstate for a given value of $v$.  Equation \refeq{eq:adia} can be rearranged in terms of the drive rate which results in the condition $\gamma \ll \gamma^\prime$ where, for the CS model, 

\begin{equation}
\gamma^\prime = \min\limits_{v_0 \leq v \leq v_f} \frac{\vert E_{N+2} (v) - E_{m} (v) \vert^2}{\vert \langle E_{N+2}(v) \vert \hat{\sigma}_x \vert E_{m}(v) \rangle\vert}, \hspace{5pt} \text{for} \hspace{5pt} m \neq N+2,
\label{eq:gammap}
\end{equation}
where $\vert E_{N+2} (v) \rangle$ is the target BC state in the topological phase and we have used the fact that $\dot{\hat{H}} = -\gamma \hat{\sigma}_x$.
In Figure \ref{fig:fid} (a), the energy spectrum as a function of $v$ is shown, with the red and black curves representing the energies of the bound and bulk states, respectively.  The energy of the target state is the top red curve in the image.  
Although the gap between the two $E = 0$ eigenstates closes, each eigenstate is an opposite parity superposition of the two bound states.  Because the initial state in Eq.\ \eqref{eq:init} has even parity and the driven operator $\hat{\sigma}_x$ does not couple states of opposite parity, $\gamma^\prime$ stays nonzero throughout the driving period for any finite system size.  We find that $\gamma^\prime$ is a minimum near the critical point ($v = w$) and corresponds to the adjacent even parity eigenstate ($m = N+4$).  Numerically, we find $\gamma^\prime \approx 22.35 N^{-1.33}$  (see Appendix \ref{app:FSS}) suggesting a finite size scaling exponent of 4/3.  For the system sizes used in this work, $100\leq N \leq 200$, this corresponds to $1.9\times 10^{-2} \leq \gamma^\prime \leq 4.8 \times 10^{-2}$.  Throughout the paper we use a drive rate of $\gamma = 1.2 \times 10^{-3}$ which satisfies the adiabaticity condition.

With adiabaticity satisfied, the fidelity of the final state is determined by the fidelity of the initial state.  In the topologically trivial phase, the identical spins' component of the eventual bound states takes the form of Eq.\ \eqref{eq:EBS} with $n_b = 0$ and $\sigma = \sqrt{\frac{N}{4 \sqrt{1-\frac{w}{v}}}}$ (see Appendix \ref{app:trivfluc} for calculation), so when $v \gg w$, the standard deviation is $\sigma \approx \sqrt{N/4}$ and they are approximately equal to the initial coherent state in Eq.\ \eqref{eq:coh}.  Thus, the quality of the final BC state improves as the system is initialized deeper within the topologically trivial phase. This trend is illustrated in Figure \ref{fig:fid} (b), where the final fidelity is displayed to asymptotically approach unity as $v_0 \to \infty$.  Numerical results of the finite size scaling of the initial fidelity can be found in Appendix \ref{app:FSS}. 

We now briefly discuss the effects of chiral symmetry breaking terms such as $v_z \hat{\sigma}_z$ and $2w_z \hat{\sigma}_z\hat{S}_z/N$, and estimate the  allowable strengths of these perturbations for the BC state to remain generable.  Both terms shift the energies of the even and odd parity superpositions of the bound states away from $E=0$, but in different ways.  The first term breaks parity symmetry by destroying the invariance of the Hamiltonian under the combined transformation $\hat{\sigma}_z \to - \hat{\sigma}_z$ and $\hat{S}_z \to -\hat{S}_z$, leading to an energy splitting of $\pm E_{v_z} = \pm v_z$ to leading order for small $v_z$.  The second term preserves parity symmetry, but the factor of $\hat{S}_z$ introduces a higher energy cost for eigenstates with components farther away from $n =  0$, i.e., BC states whose superposition has larger separation.  This term increases the energies of both parity eigenstates by the same amount, and for small $w_z$ we estimate the shift to be $E_{w_z} = 2w_z n_b/N$, where $n_b =  N\sqrt{1-(v/w)^2}/2$ is the mean-field magnitude of the bound state location calculated in the previous subsection.

The chiral symmetry breaking perturbations will reduce the energy gap, $E_\mathrm{gap}$, between the bound and bulk states by a small amount, which in turn slightly decreases $\gamma^\prime$.  This reduction can be compensated by slowing the drive rate, and as long as the gap remains open, the BC state can still be generated.   Therefore, we require the shifts $E_{v_z}$ and $E_{w_z}$ to satisfy $E_{v_z}, E_{w_z} \ll E_\mathrm{gap}$.  To estimate $E_\mathrm{gap}$, we first note that in the topological phase ($v/w <1$) the energy spectrum is approximately independent of $v$, as seen in Fig.\ \ref{fig:fid}(a).  This allows us to set $v = 0$, leaving only the spin-exchange term in $\hat{H}$.  In this limit, we are able to diagonalize $\hat{H}$ in the subspace spanned by $\vert n, \uparrow \rangle$ and $\vert n+1, \downarrow\rangle $ for each $n$ giving energies $E_{n,\pm} = \pm 2w\sqrt{\left (N/2-n \right )\left (N/2+n+1 \right )}/N$.  The energy gap is then $E_\mathrm{gap} \approx E_{N/2-1,+}- E_{N/2,+} = 2w/\sqrt{N}$.  Thus, the BC state remains generable provided $v_z \ll 2w/\sqrt{N}$ and $w_z \ll 2w/\sqrt{N\left[1-(v/w)^2 \right ]}$.


%

%

\subsection{Replacing the identical spins with a single bosonic mode}

Coherent cat states of light have been achieved using a variety of methods, so it is important to determine whether it is possible to do the same using our protocol.  To this end, we replace the identical spin operators in Eq.\ \eqref{eq:ham} with operators of a single bosonic mode, $2\hat{S}_-/N \to \hat{a}$ and $2\hat{S}_+/N \to \hat{a}^\dagger$, where $\hat{a}^\dagger$ ($\hat{a}$) is the creation (annihilation) operator of the mode obeying the usual commutation relation $[\hat{a},\hat{a}^\dagger] = 1$.  The Hamiltonian becomes

\begin{equation}
\hat{H} \to \hat{H}_a = v\hat{\sigma}_x + w \left (\hat{a}^\dagger\hat{\sigma}_- + \hat{a}\hat{\sigma}_+ \right ).
\end{equation}
where the term proportional to $w$ is now in the form of a Jaynes-Cummings interaction which is found in the system of a qubit interacting with a single mode of an optical cavity and describes the absorption and stimulated emission of photons by the qubit.  Like in the case of the identical spins, we assume that if there are any bound states in the Fock space of the bosonic mode, then they are eigenstates of $\hat{\sigma}_z$

\begin{align}\tilde{\chi}_{\uparrow}(x) = A_\uparrow(x) 
 \begin{bmatrix}
         1 \\
	0
 \end{bmatrix} \nonumber \\
\tilde{\chi}_{\downarrow}(x) = A_\downarrow(x)
 \begin{bmatrix}
         0 \\
	1
 \end{bmatrix} 
\end{align}
and they have zero energy $\hat{H}_a \tilde{\chi}_{\uparrow(\downarrow)}(x) = 0$.  In the $x$-representation of the mode operators, the energy equation becomes

\begin{equation}
\left [\left ( v +\frac{w}{\sqrt{2}}x \right ) \hat{\sigma}_x + i\frac{w}{\sqrt{2}} \frac{\partial}{\partial x} \hat{\sigma}_y \right ]\tilde{\chi}_{\uparrow(\downarrow)} (x) =0 \nonumber \\
\label{eq:e0}
\end{equation}
which has solutions of the form

\begin{eqnarray}
A_\uparrow(x) &=& C_\uparrow  e^{\frac{1}{2} (x+ \frac{\sqrt{2}v}{w})^2} \nonumber \\
A_\downarrow(x) &=& C_\downarrow  e^{-\frac{1}{2} (x+ \frac{\sqrt{2}v}{w})^2}.
\end{eqnarray}
Only the second solution is normalizable with $C_\downarrow = \pi^{-1/4}$ and is simply the $x$-representation of the coherent state $\vert \alpha = - v/w \rangle = e^{-\frac{1}{2}\vert \frac{v}{w} \vert^2} \sum_{m_a} \frac{(-v/w)^{m_a}}{\sqrt{m_a!}} \vert m_a\rangle$ where $\hat{a}^\dagger\hat{a}\vert m_a \rangle = m_a\vert m_a\rangle$.  This means there is only a single bound state and it exists in the spin-$\downarrow$ subspace.  The BC state protocol relies on there being two bound states, each one attached to different qubit states, that can used to  split the initial state into two distinguishable parts.  Therefore, we conclude that the protocol will not work if the identical spins are replaced with a single bosonic mode.

\begin{figure}[t]
\centering
\includegraphics[width=\columnwidth]{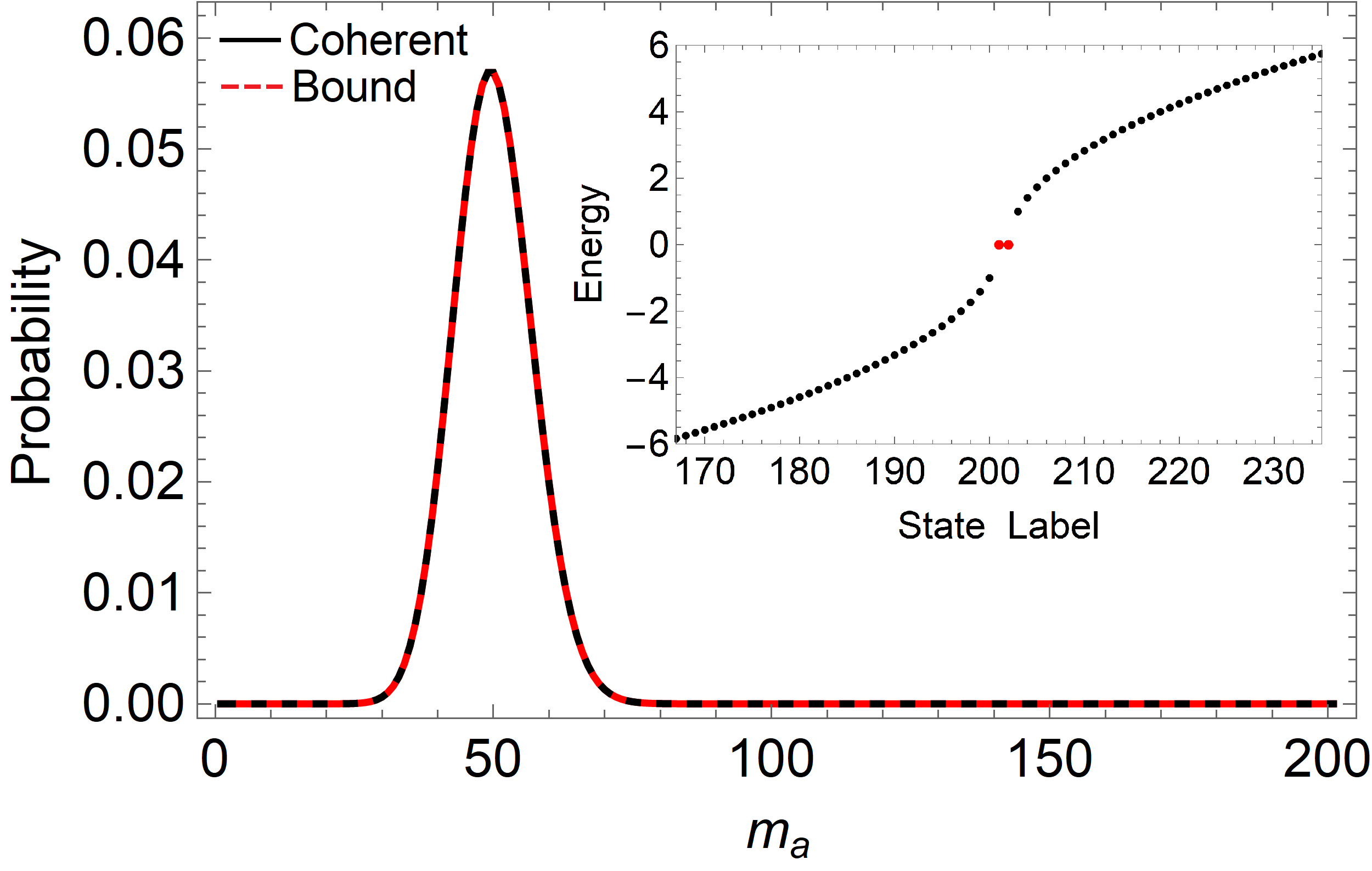}
\caption{Probability distribution of the bound state (red, dashed) of $\hat{H}_a$ in Eq.\ \eqref{eq:hama} alongside a coherent state (black, solid).  The number of states of the bosonic mode is truncated at $N_a = 201$ and the parameters are $v = 7$ and $w = 1$, so both states are centered at $m_a = 49$.   Inset: Zoom in of the spectrum of $\hat{H}_a$ around the bound states (red) at $E=0$.}
\label{fig:cohbound}
\end{figure}

The presence of a single bound state can be explained by the bulk-boundary correspondence, which asserts that a topological invariant, such as the winding number in 1D, counts the number of bound states at each boundary separating regions of different topology. In the case of $N$ identical spins coupled to a qubit, there are two bound states because the Fock space has two boundaries at $n_\mathrm{edge} = \pm N/2$. However, for a single bosonic mode, the Fock space is only bounded from below by the vacuum, meaning there is only one edge and, thus, a single bound state. This bulk-boundary correspondence can be verified by calculating the mean-field winding number, where the mode operators are replaced with their coherent state expectation values $\langle \alpha \vert \hat{a} \vert \alpha \rangle = \vert \alpha \vert e^{-i\beta}$ leading to the expression

\begin{equation}
H_a = v\hat{\sigma}_x + w |\alpha| \left ( \cos\beta \hat{\sigma}_x + \sin\beta \hat{\sigma}_y \right ).
\label{eq:hama}
\end{equation}
Comparing this equation with Eqns.\ \eqref{eq:bloch} and \eqref{eq:comp}, we see they are the same when the transformations $\sin\theta \to |\alpha|$ and $\phi \to \beta$ are made.  Applying these transformations to the calculation of the winding number in Eq.\ \eqref{eq:wind} gives 

\begin{equation}
W_a(\alpha) = \frac{1}{2}\left [1+\mathrm{sgn} \left (w|\alpha| - v \right )\right ].
\label{eq:aW}
\end{equation}
Equation \ref{eq:aW} shows that when both $w$ and $v$ are nonzero, the winding number changes once at $\vert \alpha \vert = v/w$, which corresponds to the location of the bound state. Figure \ref{fig:cohbound} displays the Fock space probability distributions of the bound state, obtained numerically by diagonalizing Eq.\ \eqref{eq:hama} (red dashed curve), alongside the coherent state centered at $m_a = \vert \alpha \vert^2 = \vert v/w\vert^2$ (black solid curve).  The inset shows the energy spectrum of $\hat{H}_a$, highlighting \textit{two} bound states at $E = 0$ in red. The first is the bound state featured in the main figure, while the second arises due to the truncation of the bosonic mode’s Hilbert space at $N_a = 201$, which introduces an artificial boundary. According to the bulk-boundary correspondence, this boundary supports an additional bound state at the opposite edge. This degeneracy can be lifted, even in the truncated Hilbert space, by adding a perturbative term to $\hat{H}_a$ proportional to the mode energy, $\hat{a}^\dagger\hat{a}$. The perturbation must remain small enough to avoid closing the energy gap between the bound state and the bulk, which would otherwise destroy the true bound state.

\subsection{Random noise in the driving}

In general, the effect of the interactions between a system and an environment causes the destruction of quantum superpositions through a process called decoherence.  Decoherence can come from many sources such as unwanted noise in laboratory equipment or from  measurement processes.  The stability of the bound states originates from the chiral symmetry which is maintained throughout the driving process, however, we can consider the case where the magnetic field driving the CS is noisy.  In general, all components of the magnetic field can introduce noise, however, the $z$-component is the most important because it breaks chiral symmetry in the Hamiltonian, so we consider the introduction of the small noisy term $\xi(t)\hat{\sigma}_z$.  Assuming white noise such that $\langle \xi(t_1)\xi(t_2)\rangle \propto \delta(t_1-t_2)$ and that the Born approximation can be applied, the density matrix evolves according to the master equation (of Lindblad form)


\begin{equation}
\frac{d\rho(t)}{dt} = i \left [ \rho(t),\hat{H}(t) \right ] + \mathcal{D}[ \hat{\sigma}_z  ]\rho(t)
\label{eq:ME}
\end{equation}
where the first term describes the coherent evolution of the density matrix and 

\begin{equation}
\mathcal{D}[ \hat{A}]\rho = \frac{D}{2} \left (2\hat{A}\rho\hat{A} - \hat{A}^2\rho - \rho\hat{A}^2 \right )
\end{equation}
is the dissipator.  The parameter $D$ is proportional to the square of the $z$-component of the magnetic field which is assumed to be small.  We note that a similar master equation describes  weak continuous  measurements of the $\hat{\sigma}_z$ eigenstates where the result of each measurement is not read out.  In that case, the source of  $D$ is the back-action of the measurement of $\hat{\sigma}_z$ on the system where the more accurate the measurement, the larger it is.  Equation \eqref{eq:ME} describes a type of decoherence called dephasing which results in the scrambling of the phases of the wave function or in the case of the density matrix, decay of coherences which are contained in its off-diagonal elements.

\begin{figure}[t]
\centering
\includegraphics[scale=0.16]{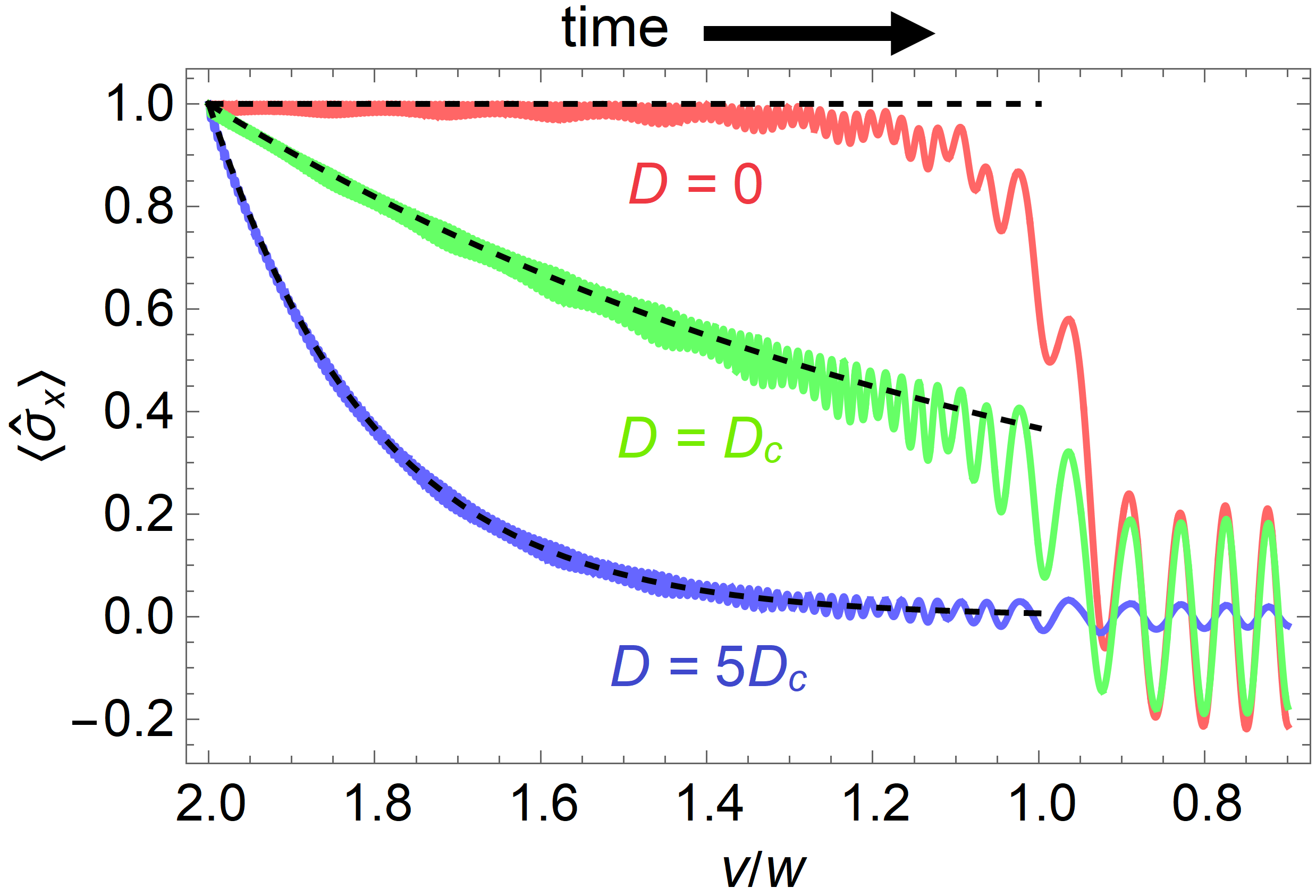}
\caption{Coherence of the CS as a function of $v$ for different decoherence strengths as the system is driven through the topological phase transition point at $v = w$.  The parameter values are $w = 1$ and $N = 100$.}
\label{fig:sigx}
\end{figure}

In our protocol, the time needed to form a BC state is $t_\mathrm{BC} = t_1 + t_2$ where $t_1 = (v_0-w)/\gamma$ is the time it takes to reach the topological transition point at $v = w$ and $t_2$ is the time from $t_1$ when the wave function first splits into a macroscopic superposition.  We define the macroscopic superposition as being formed when the separation between the peaks of the spin-$\uparrow$ and spin-$\downarrow$ components of the wave function become equal to one of their widths.  Using the mean and standard deviation from the semiclassical bound state wave function in Eq.\ \eqref{eq:EBS}, the splitting occurs when $2n_b(t_2) = \sigma(t_2)$ where the time dependence comes from the driving of the spin-flip energy $v(t) = v_0 - \gamma t$.  Solving for $t_2$ gives $t_2 \approx w/\left(128^{1/3} N^{2/3} \gamma\right )$.  Due to the $N$ dependence of $t_2$, for sufficiently large $N$, $t_1$ becomes the dominant term resulting in $t_\mathrm{BC} \approx (v_0-w)/\gamma$.  

In order for a BC state to form, the condition  $t_\mathrm{BC} \ll t_D$ must be satisfied, where $t_D$ is the decoherence time.  To simplify the analysis, we assume that throughout the driving process in the topologically trivial phase the initial state has high fidelity with the target energy eigenstate and that adiabaticity is maintained.  This means that there is negligible coherent dynamics, so the first term in Eq.\ \eqref{eq:ME} can be ignored.  Now the identical spins can be traced out, $\rho_\mathrm{CS}(t) = \mathrm{Tr}_S\left[\rho(t)\right]$, and we are left with a master equation in terms of the CS only

\begin{equation}
\frac{d\rho_\mathrm{CS}(t)}{dt} \approx D \left [\hat{\sigma}_z \rho_\mathrm{CS}(t) \hat{\sigma}_z - \rho_\mathrm{CS}(t) \right ].
\end{equation}   
With the initial state being $\vert \psi(0) \rangle$ from Eq.\ \eqref{eq:init}, the master equation can be solved for each component of  $\rho_\mathrm{CS}(t)$ giving

\begin{equation}
\rho_\mathrm{CS}(t) \approx \frac{1}{2}\begin{pmatrix}
1 & e^{-t/t_D} \\
e^{-t/t_D} & 1
\end{pmatrix}
\label{eq:pred}
\end{equation}
where the decoherence time is $t_D = \frac{1}{2D}$.  Therefore, there is a crossover region around $D_c = \frac{\gamma}{2 (v_0 - w)}$ where for $D \gg D_c$, the system will decohere before the BC state has a chance to form.  

To confirm this analysis we plot the CS coherence, $\langle \hat{\sigma}_x (t) \rangle  = \mathrm{Tr}\left [\hat{\sigma}_x \rho(t) \right ]$, as a function of $v$ (time) for different values of $D$ in Figure \ref{fig:sigx}.   In the absence of decoherence (red), $\langle \hat{\sigma}_x (t) \rangle$ remains close to unity, while significant decoherence (blue) results in exponential decay in the topologically trivial region ($v > 1$).  Each of the three curves closely follows the predicted result of $\langle\hat{\sigma}_x (t)\rangle= e^{-t/t_D}$ (dashed, black curves) from Eq.\ \eqref{eq:pred}.  The minor oscillations observed in the trivial region stem from the initial state in Eq.\ \eqref{eq:init} having imperfect overlap with the target energy eigenstate [see Eq.\ \eqref{eq:TT}].  Due to not being completely adiabatic, the amplitudes of the small oscillations become amplified in the topologically nontrivial region ($v<1$), however, they still oscillate around the predicted value of zero. The abrupt drop in the nontrivial region signifies the formation of the BC state, as can be explicitly seen by using the expressions for the bound states in Eq.\ \eqref{eq:EBS} to calculate the expectation value, resulting in $\langle \hat{\sigma}_x (t)\rangle_B = e^{-\frac{n_b(t)^2}{2\sigma(t)^2}}$. As the driving shifts the states to larger $n_b$, the coherence of the CS exponentially decays.

\section{Conclusions and Discussion}

We have demonstrated that in a CS model obeying chiral symmetry, a topological phase transition takes place. In the topologically nontrivial phase, a pair of protected bound states emerges within the Fock space of the identical spins surrounding the CS. Each member of this pair is coupled to a different state of the CS. Notably, these bound states exhibit a unique characteristic: their positions in Fock space can be adjusted by tuning parameters in the system, distinguishing them from typical bound states found in many condensed matter systems. This distinction stems from the spherical geometry of the Hilbert space of identical spins, which differs from the circular and toroidal geometry found in the quasi-momentum of 1D and 2D periodic lattices, respectively.

The mobility of the bound states is pivotal in the protocol employed for generating BC states. Our approach involves having a large overlap between an initial spin coherent state and a target eigenstate of the Hamiltonian which will become the protected bound states, then adiabatically driving the system through the topological phase transition to induce the splitting of the bound states into a macroscopic superposition within Fock space.  We achieve the large overlap by initializing the system in the topologically trivial phase where the bound states have the same form as the spin coherent state.  To ensure an equal overlap with the two bound states, we initialize the CS in an equal superposition of its two states.  The splitting is achieved by adiabatically driving the system through the transition point into the topologically nontrivial phase where the bound states become mobile.  The driving results in the CS spin-up part of the state moving in one direction in Fock space  while the spin-down part moves in the other direction.  When the system is driven deep enough into the nontrivial phase, the two parts of the state  have exponentially small overlap and a BC state is formed.


Although the version of the CS model we have analyzed is quite simple, it can be expanded to be more general because the main criterion needed to generate a BC state is the presence of mobile topologically protected bound states.  This requirement means that the Hamiltonian must obey chiral symmetry which leaves room for additional terms of the form $\hat{\sigma}_x f(\hat{S}_x, \hat{S}_y, \hat{S}_z)$ and $\hat{\sigma}_y g(\hat{S}_x, \hat{S}_y, \hat{S}_z)$ where $f$ and $g$ denote general functions of the identical spins' operators.  The zero energy eigenstates (there may be many pairs of such states depending on the functions) will also be eigenstates of the chiral symmetry operator, $\hat{\sigma}_z$, and will serve as the target BC state(s) in similar driving protocols to the one considered in this paper.  In experiments, it may be impossible to neglect chiral symmetry breaking terms.  We mentioned two such terms that may appear in the CS model which were $v_z \hat{\sigma}_z$ and $2w_z \hat{\sigma}_z \hat{S}_z/N$.  We considered them as perturbations, but it may be possible to mitigate their effects at arbitrary strengths.  Separately they destroy the bound states, however, together it may be possible to tune the parameters of the system such that their contributions cancel for a given bound state, $v_z \hat{\sigma}_z + w_z \hat{\sigma}_z \langle \hat{S}_z \rangle_B = 0$.  Although further analysis is necessary to explore this possibility, it could potentially offer a workaround for certain persistent symmetry breaking terms.  Finally, throughout the paper we assume that all surrounding spins couple equally to the central spin.  We remark that both chiral and parity symmetries survive if this condition is relaxed, so each spin sector still hosts BC states.  If the variation in coupling strengths is small enough that the system’s dynamics remain mostly confined to the spin-$N/2$ sector, the BC state can still be generated.

\acknowledgments 
D.H.J.O. acknowledges the support of the Natural Sciences and Engineering Research Council of Canada through Discovery Grant No.\ RGPIN-2017-06605.

\appendix

\section{Finite Size Scaling}
\label{app:FSS}
\begin{figure}[t]
\centering
\includegraphics[width=\columnwidth]{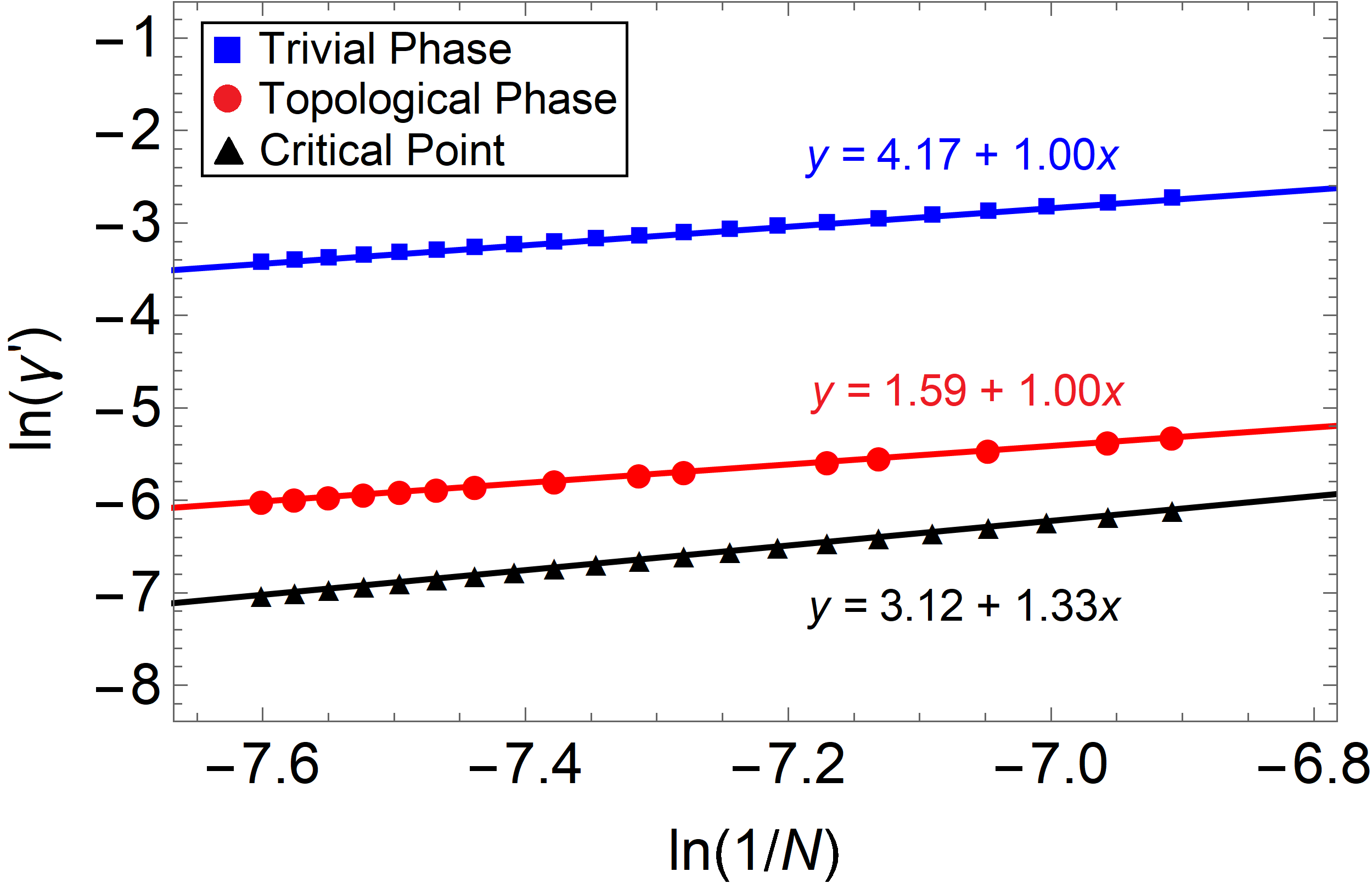}
\caption{Plot of $\ln(\gamma^\prime)$ as a function of $\ln(1/N)$.  The data corresponds to $v=0.1w$ (topological phase, red circles), $v =  w$ (critical point, black triangles) and $v = 2w$ (trivial phase, blue squares) while the equations are best fit lines.  The range of system sizes is $1000 \leq N \leq 2000$ and $w=1$.}
\label{fig:FSSG}
\end{figure}

Here, we present calculations of the finite size scaling of both $\gamma^\prime$ and the fidelity between the initial coherent state and the target eigenenergy state which becomes the BC state in the topological phase.  We relax the minimization condition over the values of $v$ in Eq.\ \eqref{eq:gammap}, so that we can calculate the finite size scaling of $\gamma^\prime$ in the trivial phase, the topological phase and at the critical point.  Figure \ref{fig:FSSG} shows a plot of $\ln(\gamma^\prime)$ as a function of $\ln(1/N)$ for $v=0.1w$ (topological phase, red circles), $v =  w$ (critical point, black triangles) and $v = 2w$ (trivial phase, blue squares).  The best fit lines displayed in the figure indicate scaling exponents consistent with $N^{-1}$ in both the topological and trivial phases, and $N^{-4/3}$ at the critical point.  From the image we also see that $\gamma^\prime$ is the smallest at the critical point which is expected as the energy gap is smallest there.

\begin{figure}[t]
\centering
\includegraphics[width=\columnwidth]{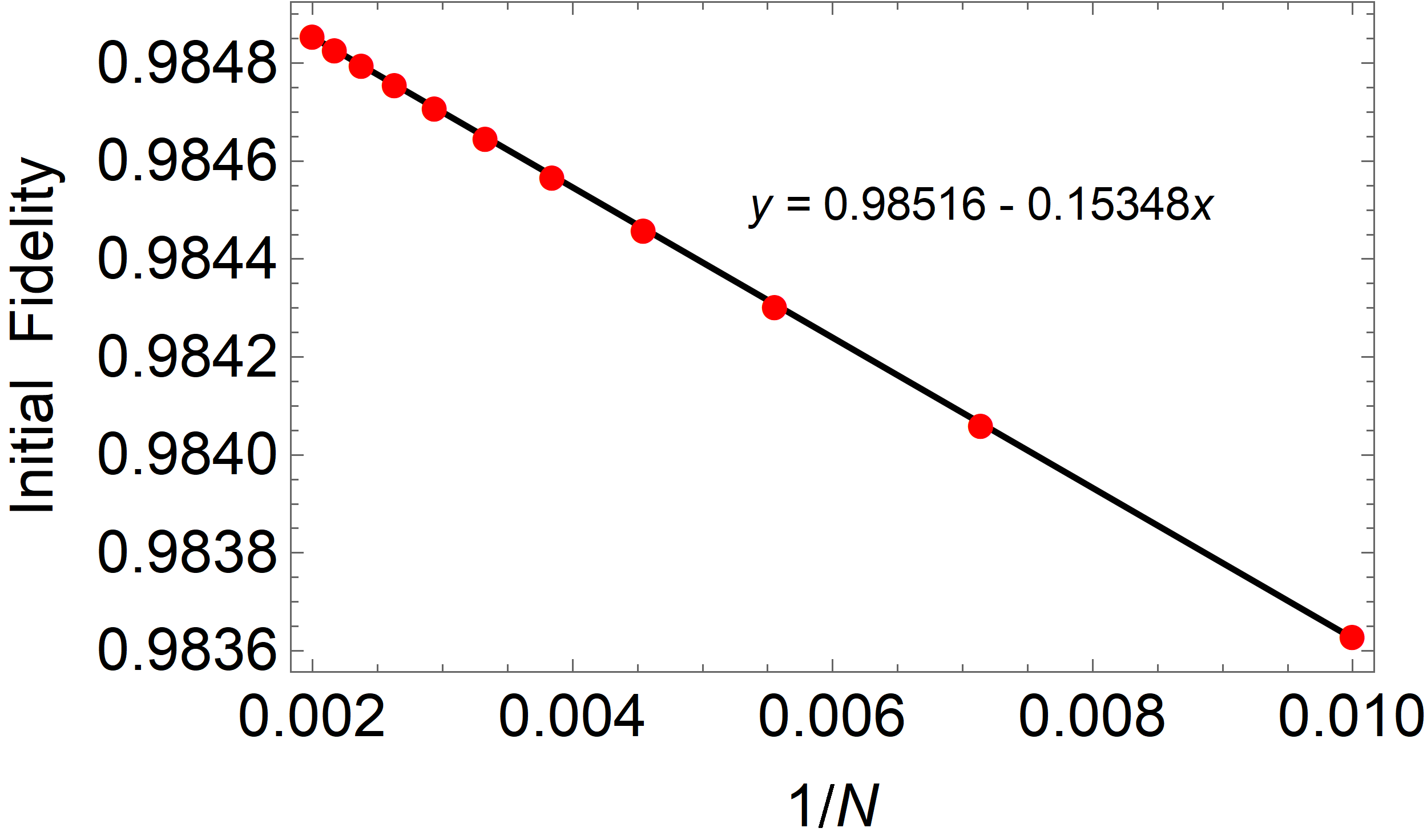}
\caption{Fidelity of the initial coherent state in Eq.\ \eqref{eq:init} and the target state as a function of $1/N$.  The range of system sizes is $100 \leq N \leq 500$ and the parameters are $v_0 = 2w$ and $w=1$.}
\label{fig:initFSS}
\end{figure}

Figure \ref{fig:initFSS} shows the fidelity between the initial coherent state and the target state as a function of $1/N$.  The best fit line takes the form $y = c_0 + c_1 x$, indicating that the system size dependent term scales as $N^{-1}$.  Although we have calculated both constants numerically, as shown in the image, $c_0$ can be obtained analytically by evaluating the fidelity between the expressions for the initial coherent state and the target state in Eqns.\ \eqref{eq:coh} and \eqref{eq:TT}, respectively.  This gives 

\begin{equation}
c_0 = \frac{2 \left ( 1-\frac{w}{v}\right )^{1/4}}{1+\sqrt{1-\frac{w}{v}}}
\end{equation}
and even for moderate initial values of $v$ such as $v_0 = 2w$, we get $c_0 = 0.98517$, which agrees well with the numerical result shown in the image. 

\section{Target eigenstate wave functions in the topologically trivial phase}
\label{app:trivfluc}

We start by deriving an expression for the identical spins' component of the bound state in the topologically trivial phase by diagonalizing Eq.\ \eqref{eq:bloch} giving

\begin{equation}
E_\pm = \pm \sqrt{v^2+w^2\left ( 1-\frac{4n^2}{N^2} \right )+2vw\cos\phi \sqrt{1-\frac{4n^2}{N^2}}}
\end{equation}
where we have used the relation $n = \frac{N}{2} \cos\theta$.  The overall sign signifies whether the CS is in an even ($+$) or odd ($-$) superposition and corresponds to the top and bottom energy bands, respectively, in Figure \ref{fig:fid} (a).   In the trivial phase, the even parity bound state is the ground state of the top band which we know is centered at $(n,\phi) = (0,\pi)$, so we expand the energy around those points keeping up to quadratic order in both variables

\begin{equation}
E_+ \approx  \frac{vw}{2(v-w)} \left (\phi - \pi \right )^2 + \frac{2w}{N^2} n^2.
\end{equation}
In the semiclassical theory, $n$ and $\phi$ are conjugate variables obeying the commutation relation $[n,\phi] =- i$, so we can make the transformation $\phi \to i\frac{d}{dn}$ resulting in

\begin{equation}
E_+ \approx  \frac{vw}{2(v-w)} \left (i\frac{d}{dn} - \pi \right )^2 + \frac{2w}{N^2} n^2
\end{equation}
where the energy describes an harmonic oscillator (with momentum shifted by $\pi$) whose mass and frequency are $ \frac{(v-w)}{vw}$ and $ \frac{2w}{N}\sqrt{\frac{v}{v-w}}$, respectively.  The ground state of the system is 

\begin{equation}
B_0(n) \approx \left(\frac{2}{\pi N}\sqrt{1-\frac{w}{v}}  \right )^{1/4} e^{i\pi n} e^{-\sqrt{1-\frac{w}{v}} \frac{n^2}{N}}
\label{eq:TT}
\end{equation}
where the phase factor $e^{i \pi n}$ comes from the $\pi$ shift of $\phi$.  One can see that as $w/v \to 0$, $B_0(n)$ approaches the spin coherent state in Eq.\ \eqref{eq:coh}.  Therefore, for larger $v_0$ the spin coherent state has a larger initial overlap with Eq.\ \eqref{eq:TT} and if the system is evolved adiabatically, the final overlap will also be larger as shown in Figure \ref{fig:fid} (b).

\end{document}